\documentclass[aps,pra,preprint,showpacs]{revtex4}
\usepackage{graphicx}
\usepackage{bm}
\usepackage{subfigure}
\usepackage{amsmath}
\usepackage[utf8]{inputenc}

\newcommand{\epl}{Europhys. Lett.\ }

\newcommand{\pr}{Phys. Rev.\ }
\newcommand{\pla}{Phys. Lett. A\ }
\newcommand{\jpa}{J. Phys. A\ }
\newcommand{\jpb}{J. Phys. B\ }
\newcommand{\njp}{New J. Phys.\ }
\newcommand{\etal}{{\em et al. }}
\newcommand{\e}{\mbox{e}}

\newcommand{\UQ}{School of Mathematics and Physics, University of Queensland, Brisbane, 
QLD 4072, Australia.}
\newcommand{\Otago}{Quantum Science Otago, Department of Physics, University of Otago, Dunedin, New Zealand.}

\begin{document}

\title{A quantum correlated twin atom laser from a Bose-Hubbard system}

\author{M.~K. Olsen}
\affiliation{\UQ}

\author{A.~S. Bradley}
\affiliation{\Otago}

\date{\today}

\begin{abstract}

We propose and evaluate a method to construct a quantum correlated twin atom laser using a pumped and damped Bose-Hubbard inline trimer which can operate in a stationary regime. With pumping via a source condensate filling the middle well and damping using either an electron beam or optical means at the two end wells, we show that bipartite quantum correlations build up between the ends of the chain, and that these can be measured either in situ or in the outcoupled beams. While nothing similar to our system has yet been achieved experimentally, recent advances mean that it should be practically realisable in the near future.

\end{abstract}

\maketitle

\section{Introduction}
\label{sec:intro}

The concept of the atom laser~\cite{Auckland1995} first entered the scientific literature very close to the time of the production of the first dilute gas Bose-Einstein condensate. Wiseman and Collett proposed that dark state cooling of a thermal atomic could be used to occupy a condensed lasing mode, from which atoms could coherently outcouple. This concept was refined by Ballagh \etal with a semiclassical analysis of a two component condensate, where the trapped component was coherently coupled to an untrapped component by microwave or radio frequency transitions~\cite{RobB}. The first experimental demonstration of a pulsed atom laser was reported by Mewes \etal in 1997~\cite{Mewes}, followed four days later by the observation of interference of condensates released from a double trap~\cite{Andrews} by Andrews \etal An overview of the early state of the field was given by Ketterle in 2002~\cite{Nobel}. Since those early days, there has been continual progress, with Robins \etal having produced a more recent review of progress~\cite{ANUboser}. 

Shortly after the concept of the atom laser was developed, theoretical work began on the Bose-Hubbard model with neutral atoms~\cite{BHmodel,Jaksch,BHJoel,Nemoto}, wherein condensed modes of atoms were trapped in the lowest energy states of an optical lattice. This model has since been constructed experimentally~\cite{Oberthaler2008}, with measurements of many of the theoretically predicted mean-field effects.

 We show here how these two fields can be combined via recent  advances in the techniques of configuring optical potentials~\cite{painting,tylerpaint} and in the outcoupling of trapped atomic modes utilising either electron beams~\cite{NDC}, or optical methods~\cite{Weitenberg}, to produce a quantum correlated twin atom laser. Our model uses an inline Bose-Hubbard trimer, with coherent pumping into the middle well and outcoupling from the two end wells, and is a pumped and damped development of the atomic mode splitter described by Chianca and Olsen~\cite{BECsplit} and compared to an optical beamsplitter~\cite{splitOC}. As we will show in what follows, both the modes in the end wells and the modes outcoupled from them possess non-classical correlations such as entanglement and Einstein-Podolsky-Rosen steering. Squeezing of an atom laser output and quantum correlations in a twin atom laser have also been proposed using a single unpumped condensate as a source, using either the collisional interactions~\cite{Simon1} or interaction with two-mode squeezed light~\cite{Simon2} to cause the quantum correlations. 

Bose-Hubbard models with pumping and loss have previously been analysed~\cite{Kordas1,Kordas2,BHcav2,BHEPR2}, predicting some interesting physical effects, both in the mean fields and the quantum statistical features. Kordas \etal have also analysed triangular dimers and inline chains with dissipation at one well~\cite{Kordas3,Kordas4}, predicting some interesting phenomena. A variety of different theoretical techniques have been used to analyse these systems, some approximate and some exact. We have chosen to use the exact mapping of our system onto stochastic differential equations in the positive-P representation~\cite{P+}, which are able to be integrated numerically with no difficulties for this damped system.

\section{Physical model, Hamiltonian, and equations of motion}
\label{sec:model}

\begin{figure}[tbhp]
\includegraphics[width=0.75\columnwidth]{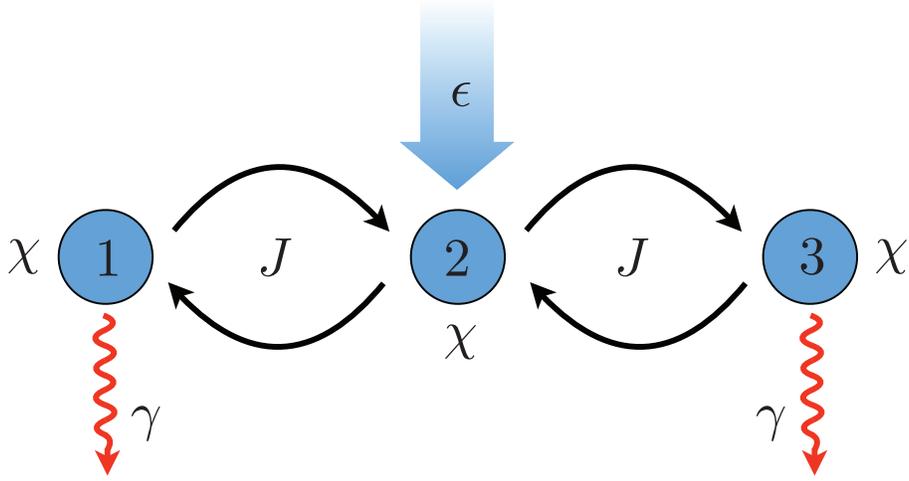}
\caption{(colour online) Schematic of the system, showing the three wells, the tunnel couplings, the pumping, and the losses.}
\label{fig:schematic}
\end{figure}

Our system is as shown in the schematic of Fig.~\ref{fig:schematic}.
In order to provide a mathematical description, we begin with the Bose-Hubbard unitary Hamiltonian for the open trimer,
\begin{equation}
{\cal H} = \hbar\chi\sum_{i=1}^{3}\hat{a}_{i}^{\dag\,2}\hat{a}_{i}^{2}-\hbar J \left[\hat{a}_{2}^{\dag}(\hat{a}_{1}+\hat{a}_{3})
+\hat{a}_{2}(\hat{a}_{1}^{\dag}+\hat{a}_{3}^{\dag}) \right],
\label{eq:genHam3line}
\end{equation}
where $\hat{a}_{i}$ is the bosonic annihilation operator for the $i$th well, $\chi$ represents the collisional nonlinearity and $J$ is the tunneling strength. The pumping into the middle well is represented by the Hamiltonian
\begin{equation}
{\cal H}_{pump} = i\hbar\left(\epsilon\hat{a}_{2}^{\dag}-\epsilon^{\ast}\hat{a}_{2}\right),
\label{eq:pump}
\end{equation}
which is of the form commonly used for the pumping of optical cavities. The basic assumption here is that this well receives atoms from a coherent condensate, represented by the c-number $\epsilon$, which is much larger than any of the modes we are investigating, so that it will not become significantly depleted over the time scales of interest.
The damping acts on the system density matrix as the Lindblad superoperator
\begin{equation}
{\cal L}\rho = \gamma\sum_{i=1,3}\left(2\hat{a}_{i}\rho\hat{a}_{i}^{\dag}-\hat{a}_{i}^{\dag}\hat{a}_{i}\rho-\rho\hat{a}_{i}^{\dag}\hat{a}_{i}\right),
\label{eq:damp}
\end{equation} 
and $\gamma$ is the coupling between the damped well and the atomic bath, which we assume to be initially unpopulated.
If the lost atoms fall under gravity, we are justified in using the Markov and Born approximations~\cite{JHMarkov,MWJack}.

Following the usual procedures~\cite{QNoise,DFW}, we map the Hamiltonian and the Lindblad term onto a Fokker-Planck equation for the positive-P function. This is then mapped onto stochastic differential equations in the It\^{o} calculus~\cite{SMCrispin},
\begin{eqnarray}
\frac{d\alpha_{1}}{dt} &=& -\left(\gamma+2i\chi\alpha_{1}^{+}\alpha_{1}\right)\alpha_{1}+iJ\alpha_{2} + \sqrt{-2i\chi\alpha_{1}^{2}}\;\eta_{1}, \nonumber \\
\frac{d\alpha_{1}^{+}}{dt} &=& -\left(\gamma-2i\chi\alpha_{1}^{+}\alpha_{1}\right)\alpha_{1}^{+}-iJ\alpha_{2}^{+} + \sqrt{2i\chi\alpha_{1}^{+\,2}}\;\eta_{2}, \nonumber \\
\frac{d\alpha_{2}}{dt} &=& \epsilon-2i\chi\alpha_{2}^{+}\alpha_{2}^{2} + iJ\left(\alpha_{1}+\alpha_{3}\right) + \sqrt{-2i\chi\alpha_{2}^{2}}\;\eta_{3}, \nonumber \\
\frac{d\alpha_{2}^{+}}{dt} &=& \epsilon^{\ast}+2i\chi\alpha_{2}^{+2}\alpha_{2} - iJ\left(\alpha_{1}^{+} +\alpha_{3}^{+}\right) +\sqrt{2i\chi\alpha_{2}^{+\,2}}\;\eta_{4}, \nonumber \\
\frac{d\alpha_{3}}{dt} &=& -\left(\gamma+2i\chi\alpha_{3}^{+}\alpha_{3}\right)\alpha_{3} + iJ\alpha_{2} + \sqrt{-2i\chi\alpha_{3}^{2}}\;\eta_{5}, \nonumber \\
\frac{d\alpha_{3}^{+}}{dt} &=& -\left(\gamma-2i\chi\alpha_{3}^{+}\alpha_{3}\right)\alpha_{3}^{+} - iJ\alpha_{2}^{+} + \sqrt{2i\chi\alpha_{3}^{+\,2}}\;\eta_{6},
\label{eq:PPequations}
\end{eqnarray}
where the $(\alpha_{j},\alpha_{j}^{+})$ are the c-number variables corresponding to the operators $(\hat{a}_{j},\hat{a}_{j}^{\dag})$ in the sense that the averages $\overline{\alpha_{j}^{m}\alpha_{k}^{+\,n}}$ converge in the limit over a large number of stochastic trajectories to the expectation values of normally-ordered operator products $\langle\hat{a}_{k}^{\dag\,n}\hat{a}_{j}^{m}\rangle$. In general, $\alpha_{i}$ and $\alpha_{i}^{+}$ are not complex conjugates, with this freedom allowing us to exactly represent quantum evolution using classical c-number variables. In the above equations, $\epsilon$ is the coherent pump amplitude from the large reservoir condensate, $\gamma$ is the loss rate at wells $1$ and $3$, and the $\eta_{j}$ are Gaussian random variables with the correlations $\overline{\eta_{j}(t)}=0$ and $\overline{\eta_{j}(t)\eta_{k}(t')}=\delta_{jk}\delta(t-t')$. These equations are solved numerically, taking averages over a large number of stochastic trajectories, of the order of $10^{6}$ for the results presented herein.

Without the random noise terms, the above collapse to three coupled semi-classical equations since the conjugate properties are then preserved.
These are useful for obtaining some semi-classical results, although in general there are no analytical solutions except in the case where $\chi=0$, with solutions
\begin{eqnarray}
\alpha_{1}^{ss} &=& \alpha_{3}^{ss} = \frac{i\epsilon}{2J} \nonumber \\
\alpha_{2}^{ss} &=& \frac{\gamma\epsilon}{2J^{2}}.
\label{eq:ssalpha}
\end{eqnarray}
 
What the above solutions do tell us is that the phases of $\alpha_{2}$ and $\alpha_{3}$ are the same, which is also obvious from the symmetry of the system. The degree of entanglement between these modes will depend on the extent to which their quantum fluctuations are correlated. To calculate this we need to proceed numerically.
For the results given in this paper, we have integrated Eq.~\ref{eq:PPequations} over at least $10^{6}$ stochastic trajectories for each parameter set, achieving good convergence of the results.
The fact that we are able to use the positive-P method at all here is due to the damping of the system, without which the equations for a Bose-Hubbard trimer in that representation are highly unstable. We have used the parameters $\gamma=1$, $J=1$ (which sets the time scale), $\chi = 10^{-2}$ and $10^{-3}$ and values for the coherent pumping amplitude, $\epsilon$, of $10$ and $10\sqrt{2}$. We found that the higher value of $\chi$ gave stronger quantum correlations, as did increasing $\epsilon$.

\begin{figure}[tbhp]
\includegraphics[width=0.75\columnwidth]{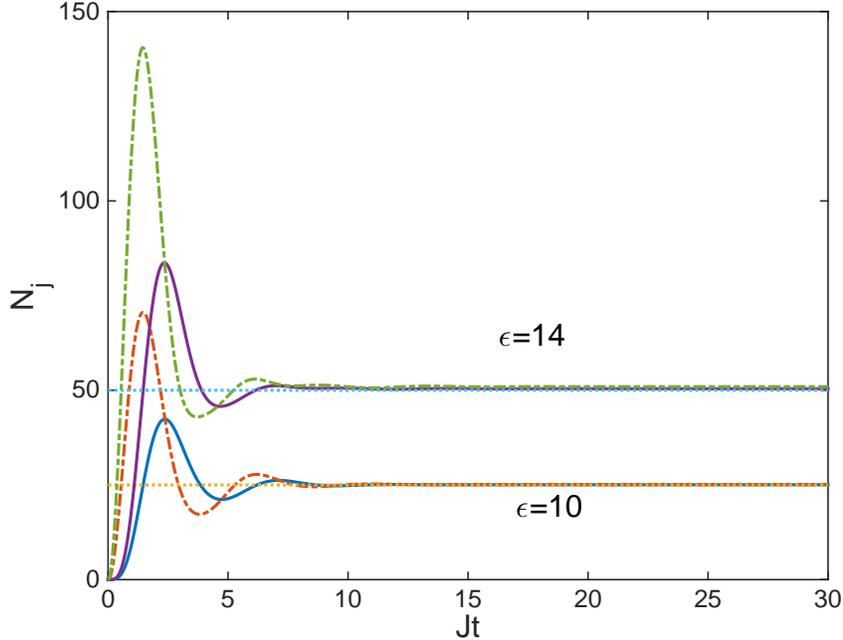}
\caption{(colour online) The populations of the wells for $\chi=10^{-3}$, $\gamma=1$, and two different pumping amplitudes. The solid lines are $N_{1}$ and $N_{3}$, which are equal, while the dash-dotted lines are $N_{2}$. The classical non-interacting steady-state values are shown by the dotted lines. $Jt$ is a dimensionless time and all quantities plotted in this and subsequent plots are dimensionless.}
\label{fig:BHAxNk3}
\end{figure}

\begin{figure}[tbhp]
\includegraphics[width=0.75\columnwidth]{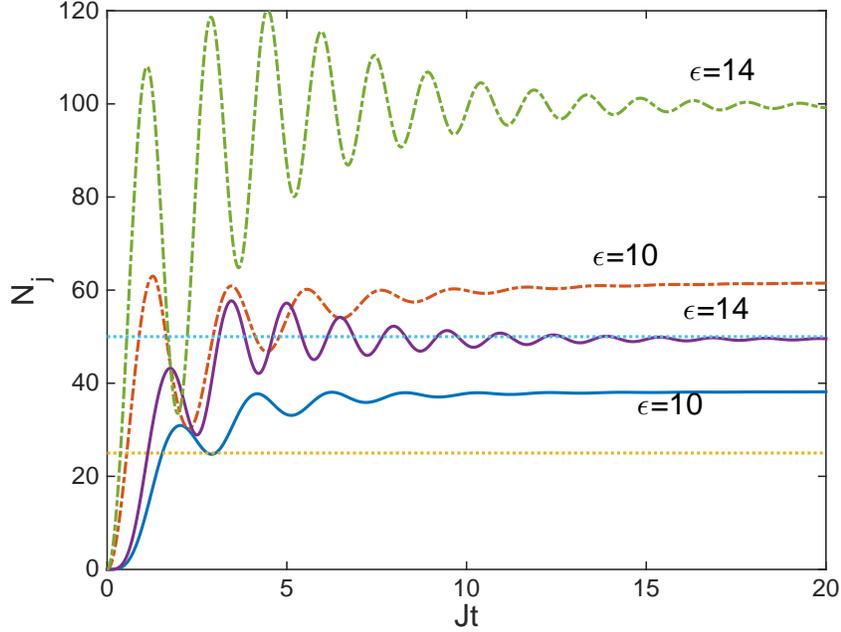}
\caption{(colour online) The populations of the wells for $\chi=10^{-2}$, $\gamma=1$, and two different pumping amplitudes. The solid lines are $N_{1}$ and $N_{3}$, which are equal, while the dash-dotted lines are $N_{2}$. The classical non-interacting steady-state values are shown by the dotted lines.}
\label{fig:BHAxNk2}
\end{figure}

In Fig.~\ref{fig:BHAxNk3} and Fig.~\ref{fig:BHAxNk2} we show the dynamics of the intrawell populations for $\chi=10^{-3}$ and $10^{-2}$, for both the pump amplitudes that we have considered. The semiclassical solutions of Eq.~\ref{eq:ssalpha} are plotted for comparison. For the lower value of $\chi$, we see that these are almost indistinguishable from the quantum solutions after the initial transients. By way of contrast, for $\chi=10^{-2}$, the non-interacting and interacting solutions are markedly different, and the populations of the middle well become different to and greater than those in the two outside wells. This effect can be understood by reference to the closed trimer considered in Refs.~\cite{BECsplit,splitOC}, in which situation the population initially in the middle well does not transfer completely to the two outer wells when $\chi=10^{-2}$ for initial populations of approximately $40$ atoms. For higher initial populations, which we see in the transients of the middle well occupations in Fig.~\ref{fig:BHAxNk2}, the closed system enters a macroscopic self-trapping type regime where tunneling is suppressed. While damping and loss change the details of this process, what we see here does serve to demonstrate that the collisional interactions are having a marked effect on the system. We also note here that the time axis for $\chi=10^{-2}$ is truncated compared to that for $10^{-3}$ because the positive-P equations ran into divergence problems after $Jt=25$. 

\section{Quantum correlations and intra-well results}
\label{sec:interest}

\subsection{Number correlations}
\label{subsec:intensity}

\begin{figure}[tbhp]
\includegraphics[width=0.75\columnwidth]{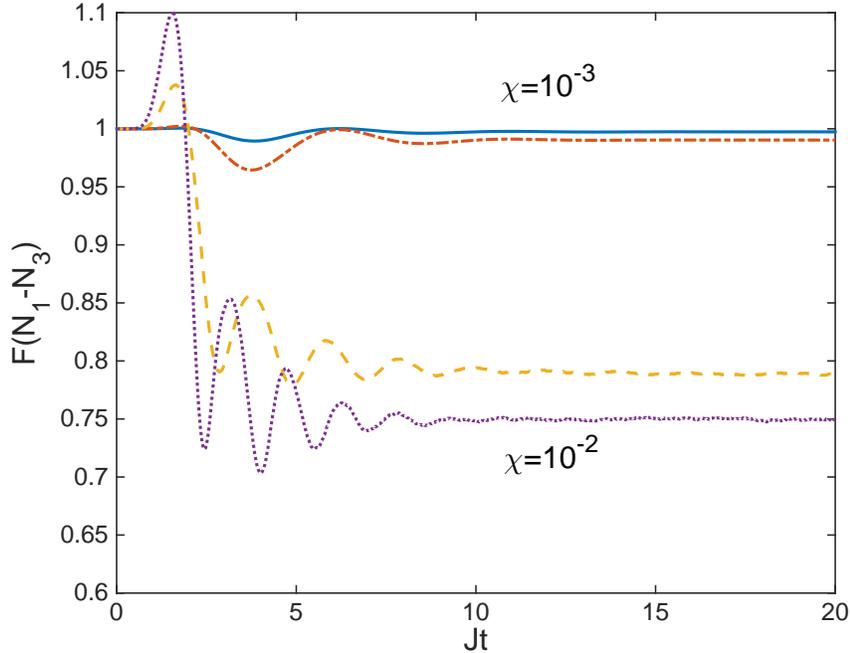}
\caption{(colour online) The normalised number difference variance for wells one and three. The solid and dashed lines are for $\epsilon=10$, while the dash-dotted and dashed lines are for $\epsilon=10\sqrt{2}$. The two upper (lower) lines are for $\chi=10^{-3}\:(10^{-2})$.}
\label{fig:Fano}
\end{figure}

There are a number of quantum statistical properties that we can investigate, including the number statistics, quadrature squeezing, and EPR-steering. We found the number statistics of each well to be less than $1\%$ different from Poissonian, which is as expected for a coherently driven and damped system. The second order equal time normalised intensity correlation functions~\cite{Roy},
\begin{equation}
g^{(2)}(N_{i}N_{j}) = \frac{\langle \hat{a}_{i}^{\dag}\hat{a}_{i} \hat{a}_{j}^{\dag}\hat{a}_{j} \rangle}{\langle \hat{a}_{i}^{\dag}\hat{a}_{i}\rangle\langle\hat{a}_{j}^{\dag}\hat{a}_{j}\rangle},
\label{eq:GRoy}
\end{equation}
between the wells are also very close to the coherent state values of one. Where we do find a difference from coherent states is in the normalised variance in the number difference between wells $1$ and $3$,
\begin{equation}
F(N_{1}-N_{3}) = \frac{V(N_{1}-N_{3})}{N_{1}+N_{3}}.
\label{eq:Fano}
\end{equation} 
which has a value of one for two independent coherent states, and a value of zero for two Fock states. 
We see from Fig.~\ref{fig:Fano} that the difference in number between the end wells is sub-Poissonian, and that increasing the nonlinearity or the pump acts to increase the intensity correlation between these two wells.
Having shown that the end wells possess what is often called relative number squeezing, we will now proceed to calculate quadrature correlations.

\subsection{Squeezing}
\label{subsec:squeeze}
        
We define the atomic quadrature operators as 
\begin{eqnarray}
\hat{X}_{j}(\theta) &= & \hat{a}_{j}\e^{-i\theta}+\hat{a}_{j}^{\dag}\e^{i\theta}, \nonumber \\
\hat{Y}_{j}(\theta) &=& \hat{X}_{j}(\theta+\pi/2),
\label{eqn:XYtheta}
\end{eqnarray}
so that the Heisenberg Uncertainty Relation is $V(\hat{X}_{j}(\theta))V(\hat{Y}_{j}(\theta))\geq 1$. It is well known in systems with a Kerr-type nonlinearity that the maximum squeezing will occur for $\theta\neq 0$~\cite{nlc,NGJoel,BHcav2}, and we find the optimal angle for our correlations by calculating them at all angles and finding the minima. Experimentally this can be done by changing the phase of the local oscillator~\cite{andyhomo}.
Squeezing in a particular quadrature exists whenever its variance, defined as $V(\hat{A})=\langle\hat{A}^{2}\rangle-\langle\hat{A}\rangle^{2}$ for any operator $\hat{A}$, is found to be less than $1$ at any angle.  In  the table below we show the minimum steady-state squeezing values and the quadrature angles at which they are found, for different values of $\chi$ and $\epsilon$. We find that $V(\hat{X}_{1})=V(\hat{X}_{3})$, and that the main factor which affects the squeezing is the collisional nonlinearity, with an increase in this giving better squeezing. Increasing the pumping amplitude has less effect, particularly on the variances of the two end modes. 

\begin{center}
 \begin{tabular}{||c || c  c || c  c ||} 
 \hline
  Variance & $\chi=$ & $10^{-3}$ & $\chi=$ &  $10^{-2}$  \\ [0.5ex] 
 \hline\hline 
 $V(\hat{X}_{1,3}),\,\epsilon=10$ & 0.93, & 130$^{o}$ & 0.69, & 15$^{o}$ \\ 
 \hline
 $V(\hat{X}_{1,3}),\,\epsilon=10\sqrt{2}$ & 0.89, & 124$^{o}$ & 0.68, & 11$^{o}$ \\ 
 \hline
 $V(\hat{X}_{2}),\,\epsilon=10$ & 0.89, & 40$^{o}$ & 0.70, & 40$^{o}$ \\ 
 \hline
 $V(\hat{X}_{2}),\,\epsilon=10\sqrt{2}$ & 0.80, & 35$^{o}$ & 0.61, & 140$^{o}$\\
 \hline
\end{tabular}
\end{center}

\subsection{Entanglement and EPR steering}
\label{subsec:EPR}
 
\begin{figure}[tbhp]
\includegraphics[width=0.75\columnwidth]{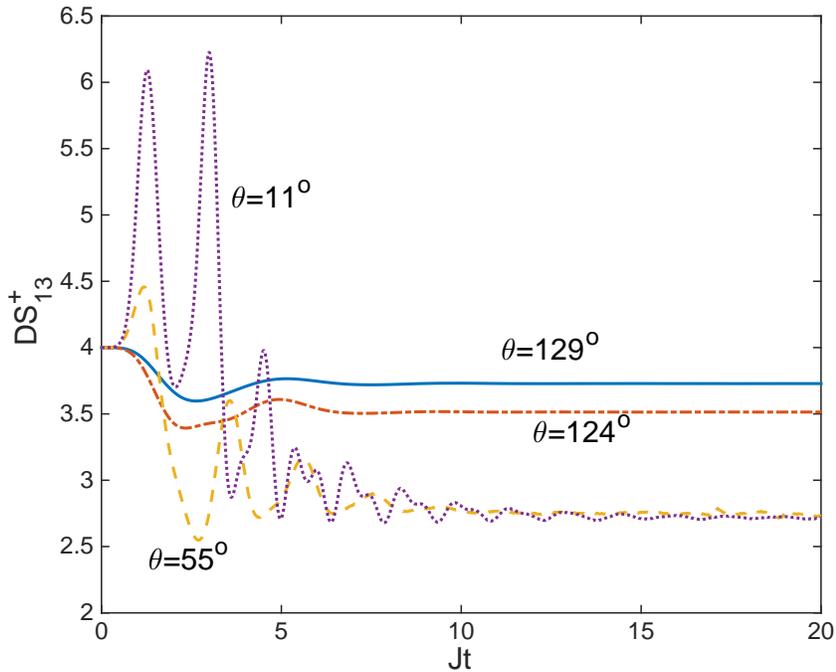}
\caption{(colour online) The values of the Duan-Simon correlation of Eq.~\ref{eq:DS}, and the angles of maximum violation of the inequality in the steady-state, for different values of the pumping and nonlinearities. The solid line is for $\chi=10^{-3}$ and $\epsilon=10$, the dash-dotted line is for $\chi=10^{-3}$ and $\epsilon=10\sqrt{2}$, the dashed-line is for $\chi=10^{-2}$ and $\epsilon=10$, and the dotted line is for $\chi=10^{-2}$ and $\epsilon=10\sqrt{2}$.}
\label{fig:BHAxDS}
\end{figure}

We now define the correlations we will investigate to detect bipartite mode entanglement. The first of these, known as the Duan-Simon inequality~\cite{Duan,Simon}, states that, for any two separable states,
\begin{equation}
V(\hat{X}_{j}+\hat{X}_{k})+V(\hat{Y}_{j}-\hat{Y}_{k}) \geq 4,
\label{eq:DS}
\end{equation}
with any violation of this inequality demonstrating the inseparability of modes $j$ and $k$. We will call this correlation function $DS^{+}_{ij}$. The violation of this inequality is necessary and sufficient to prove the inseparabilty and entanglement for Gaussian states, and sufficient for non-Gaussian states. In Fig.~\ref{fig:BHAxDS} we show the results for this inequality, demonstrating a clear violation in the steady-state. As with the quadrature squeezing, we find that the increase in collisional nonlinearity allows for a stronger violation, with increases in the pumping having somewhat less of an effect. This suggests that in any experimental measurement of these correlations, the ratio $\chi/J$ should be as large as possible.

The presence of EPR-steering~\cite{EPR,Erwin,Jonesteer} is signified by violation of the Reid inequality for the inferred variances~\cite{EPRMDR}
\begin{equation}
EPR_{ij} = V^{inf}(\hat{X}_{i})V^{inf}(\hat{Y}_{i})\geq 1,
\label{eq:eprMDR}
\end{equation}
where
\begin{equation}
V_{inf}(\hat{A}_{i}) = V(\hat{A}_{i})-\frac{[V(\hat{A}_{i},\hat{A}_{j})]^{2}}{V(\hat{A}_{j})},
\label{eq:EPRdef}
\end{equation}
and $V(\hat{A},\hat{B})=\langle\hat{A}\hat{B}\rangle-\langle\hat{A}\rangle\langle\hat{B}\rangle$.
This condition is optimal for bipartite Gaussian systems, at least sufficient for non-Gaussian systems, and also comprehensively demonstrates bipartite entanglement, as such states are a superset of the EPR states.

\begin{figure}[tbhp]
\includegraphics[width=0.75\columnwidth]{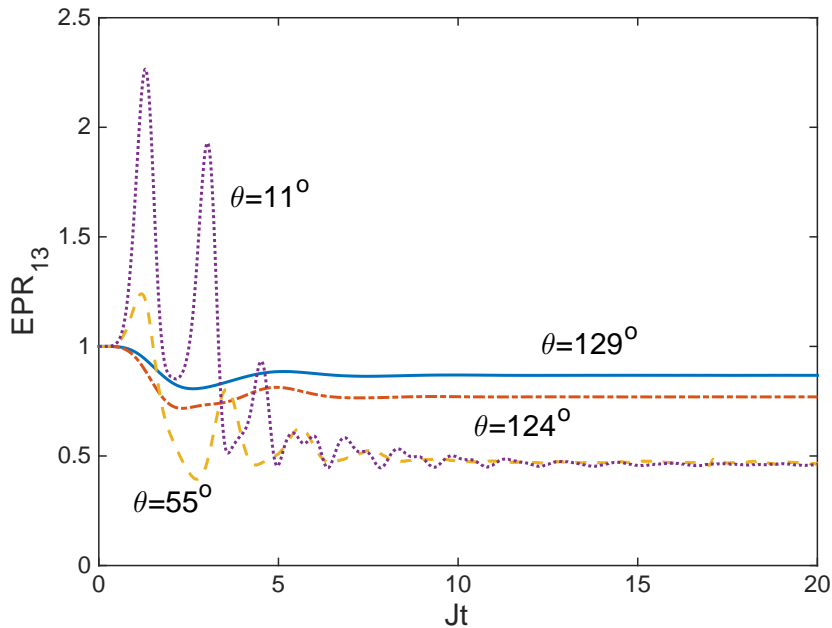}
\caption{(colour online) The values of the Reid EPR correlation of Eq.~\ref{eq:eprMDR}, and the angles of maximum violation of the inequality in the steady-state, for different values of the pumping and nonlinearities. The solid line is for $\chi=10^{-3}$ and $\epsilon=10$, the dash-dotted line is for $\chi=10^{-3}$ and $\epsilon=10\sqrt{2}$, the dashed-line is for $\chi=10^{-2}$ and $\epsilon=10$, and the dotted line is for $\chi=10^{-2}$ and $\epsilon=10\sqrt{2}$.}
\label{fig:BHAxEPR}
\end{figure}

The results for EPR-steering are shown in Fig.~\ref{fig:BHAxEPR}, where we see once again that a clearer violation is found for the larger $\chi$ value. It is of interest to note here that the maximum percentage violation of this inequality ($>50\%$) is greater than that for the Duan-Simon inequality ($<40\%$), and thus may be easier to measure experimentally. Having regard to inevitable experimental noise and any small multi-mode effects which we have ignored in our analysis, this may be crucial.

\section{Output correlations}
\label{sex:output}

Because our system is damped at the two outside wells and the atoms can fall under gravity, we may also examine any correlations in the outputs.
We may treat the system as Markovian~\cite{JHMarkov,MWJack} and therefore use the Gardiner Collett input-output relations, as long as the anharmonicity inside the trap is smaller than the damping rate, which is the case for the lower nonlinearity we have considered~\cite{GardinerCollett}. Since two out of three wells are damped rather than only one out of two~\cite{BHEPR2}, we find that the third and fourth order cumulants are insignificant and may closely approximate the system as Gaussian. In the steady state, this then allows us to treat it as an Ornstein-Uhlenbeck process~\cite{SMCrispin} and perform the standard linearised fluctuation treatment of quantum optics on the output modes. We do not consider this treatment applicable to the system with the higher collisional nonlinearity.
 
We proceed in exactly the manner followed in a previous analysis of the nonlinear Kerr coupler~\cite{nlc}, first dividing the variables into mean-value steady-state parts plus fluctuations, e.g. $\alpha_{j}=\overline{\alpha_{j}}+\delta\alpha_{j}$. The spectral matrix for the fluctuations is defined as 
\begin{equation}
S(\omega) = (A+i\omega)^{-1}D(A^{\mbox{\small{T}}}-i\omega)^{-1},
\label{eq:Sout}
\end{equation}
where
\begin{equation}
A =
\begin{bmatrix}
\gamma+2i\chi\overline{N_{1}} & 2i\chi\overline{\alpha_{1}}^{2} & -iJ & 0 &0 & 0 \\
-2i\chi\overline{\alpha_{1}^{\ast}}^{2} & \gamma-2i\chi\overline{N_{1}} & 0 & iJ & 0 & 0 \\
-iJ & 0 & 2i\chi\overline{N_{2}} & 2i\chi\overline{\alpha_{2}}^{2} & -iJ & 0 \\
0 & iJ & -2i\chi\overline{\alpha_{2}^{\ast}}^{2} & -2i\chi\overline{N_{2}} & 0 & iJ \\
0 & 0 & -iJ & 0 & \gamma+2i\chi\overline{N_{3}} & 2i\chi\overline{\alpha_{3}}^{2} \\
0 & 0 & 0 & iJ & -2i\chi\overline{\alpha_{3}^{\ast}}^{2} & \gamma-2i\chi\overline{N_{3}}
\end{bmatrix},
\label{eq:Amat}
\end{equation}
and $\overline{x}$ represents the steady-state mean of $x$, obtained from the positive-P solutions as in~\cite{Sarah}. The matrix $D$ is a $6\times 6 $ matrix with $\left[-2i\chi\overline{\alpha_{1}^{2}},2i\chi\overline{\alpha_{1}^{\ast\,2}},-2i\chi\overline{\alpha_{2}^{2}},2i\chi\overline{\alpha_{2}^{\ast\,2}},
-2i\chi\overline{\alpha_{3}^{2}},2i\chi\overline{\alpha_{3}^{\ast\,^{2}}}\right]$ on the diagonal. Because we have parametrised our system using $J=1$, the frequency $\omega$ is in units of $J$. $S(\omega)$ then gives us products such as $\delta\alpha_{i}\delta\alpha_{j}$ and  $\delta\alpha_{i}^{\ast}\delta\alpha_{j}^{\ast}$, from which we construct the output variances and covariances for modes $1$ and $3$ as
\begin{equation}
S^{out}(X_{i},X_{j}) = \delta_{ij}+\gamma \left(S_{ij}+S_{ji}\right).
\label{eq:Sout}
\end{equation}

\begin{figure}[tbhp]
\includegraphics[width=0.75\columnwidth]{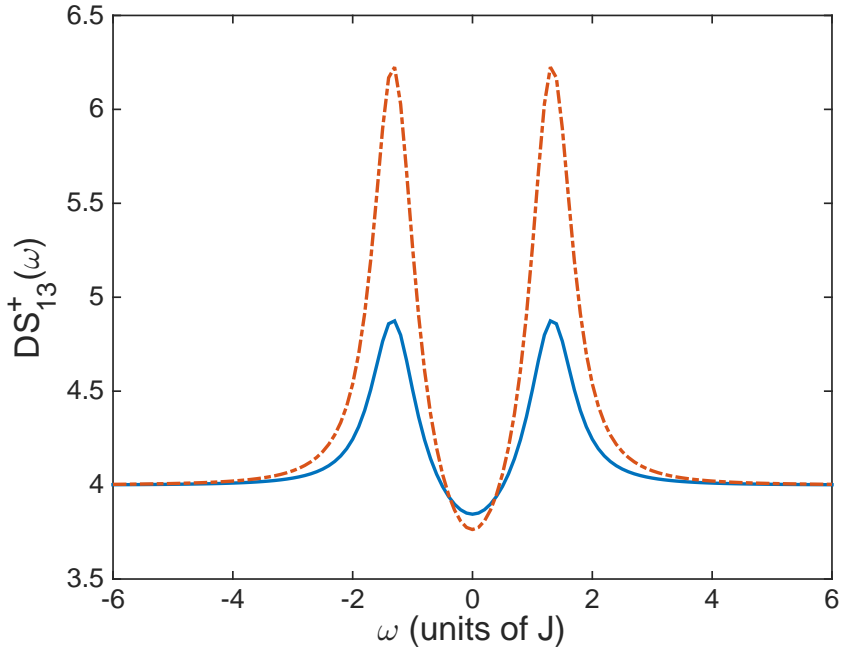}
\caption{(colour online) The spectral Duan-Simon output correlations between wells one and three for $\chi=10^{-3}$. The solid line is for $\epsilon=10$, at $\theta=129^{o}$, and the dash-dotted line is for $\epsilon=10\sqrt{2}$, at $\theta=124^{o}$.}
\label{fig:DS13out}
\end{figure}

Formally, these quantities are Fourier transforms of two-time correlation functions in the time domain. In quantum optics the frequency $\omega$ has an obvious meaning as the shift from a cavity resonance frequency in units of the inverse cavity lifetime. In our atomic case, $\omega$ represents the initial spectral energy distance from the mean steady-state mode energy in each well.

\begin{figure}[tbhp]
\includegraphics[width=0.75\columnwidth]{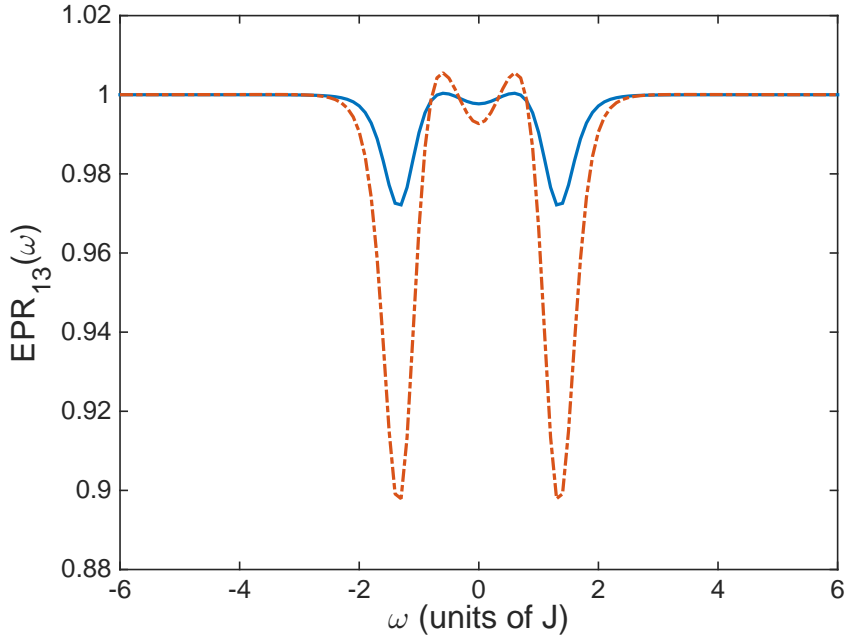}
\caption{(colour online) The spectral EPR-steering output correlations between wells one and three for $\chi=10^{-3}$. The solid line is for $\epsilon=10$, at $\theta=129^{o}$, and the dash-dotted line is for $\epsilon=10\sqrt{2}$, at $\theta=124^{o}$.}
\label{fig:EPR13out}
\end{figure}

The results of this procedure for the Duan-Simon inequality are shown in Fig.~\ref{fig:DS13out}, showing a narrow range of entanglement about the mean energy. The plots of the EPR inequality shown in Fig.\ref{fig:EPR13out} also show violations over a narrow range. The output modes from the two wells are also individually quadrature squeezed, at the same angles as shown in the plots. These results show that the two outputs possess bipartite quantum correlations. Given that the number difference statistics between the two wells are sub-Possionian, this system qualifies as a quantum correlated twin atom laser.

\section{Conclusions}
\label{sec:conclusions}

We have combined recent advances in of atomic trapping and outcoupling to outline a proposal for the realisation of a quantum correlated twin atom laser. The new techniques of potential painting along with the recent realisation of damping of individual wells via an electron beam, as well as proposals for the homodyne measurement of atomic quadrature fluctuations, allow for a system which would not have been possible in the recent past. We also note here that an equivalent system would not be possible using optical cavities since, while not all cavities need to be pumped, they inevitably undergo damping. An atomic system allows for the freedom of choice over which individual wells will be damped as well as which will be pumped. While the pumping condensate remains undepleted, our system will produce two quantum correlated steady-state beams of atoms.  

\section*{Acknowledgments}

M.K.O. was supported by the Australian Research Council under the Future Fellowships Program (Grant ID: FT100100515) and A.S.B was supported
by a Rutherford Discovery Fellowship administered by
the Royal Society of New Zealand..

\end{document}